\documentclass[preprint,showpacs,showkeys,eqsecnum,aps]{revtex4}

\begin{document}

\title{Invariant conserved currents in gravity theories: diffeomorphisms
and local gauge symmetries}

\author{Yuri N.~Obukhov}
\email{yo@thp.uni-koeln.de}
\affiliation{Institute for Theoretical Physics, University of Cologne,  
50923 K\"oln, Germany}
\altaffiliation[Also at: ]{Department of Theoretical Physics, 
Moscow State University, 117234 Moscow, Russia}
\author{Guillermo F.~Rubilar}
\email{grubilar@udec.cl}
\affiliation{Departamento de F{\'{\i}}sica, Universidad de Concepci\'on,
Casilla 160-C, Concepci\'on, Chile}

\begin{abstract}
Previously, we have developed a general method to construct invariant
conserved currents and charges in gravitational theories with Lagrangians
that are invariant under spacetime diffeomorphisms and local Lorentz
transformations.
This approach is now generalized to the case when the local Lorentz group
is replaced by an arbitrary local gauge group. The particular examples
include the Maxwell and Yang-Mills fields coupled to gravity with
Abelian and non-Abelian local internal symmetries, and the metric-affine
gravity in which the local Lorentz spacetime group is extended to the
local general linear group.
\end{abstract}


\maketitle

\section{Introduction}

As it is well known, the Noether theorem establishes a relation between
the symmetries and conservation laws of a physical model. It tells that
conserved currents arise from the invariance of the classical action when
the fields are transformed under the action of the symmetry groups. There
are two types of symmetries: the internal ones (like the Abelian $U(1)$ phase
transformations, or the non-Abelian isotopic $SU(2)$ transformations,
as well as their generalizations) that act in spaces of internal degrees
of freedom, and the external symmetries that act on the spacetime manifold
itself and on its related geometrical structures. In particular, gravity
theories are normally based on the covariance principle, which in technical
terms means that the action is invariant under spacetime diffeomorphisms.
Since diffeomorphisms are generated by vector fields, one can expect that
every vector field should give rise to conserved quantities.

In the previous paper \cite{invar}, we have proposed a general definition
of invariant conserved quantities for gravity theories with general
coordinate and local Lorentz symmetries. More exactly, we have demonstrated
that, indeed, every vector field $\xi$ on spacetime generates, in any dimension
$n$, for any Lagrangian of gravitational plus matter fields and for any (minimal
or nonminimal) type of interaction, a current ${\cal J}[\xi]$ with the following
properties: (i) the current $(n-1)$-form ${\cal J}[\xi]$ is constructed from
the Lagrangian and the generalized field momenta, (ii) it is conserved, $d
{\cal J}[\xi] = 0$, when the field equations are satisfied, (iii) ${\cal J}
[\xi] = d\Pi[\xi]$ ``on shell", (iv) the current ${\cal J}[\xi]$, the
superpotential $\Pi[\xi]$, and the conserved charge ${\cal Q}[\xi] = \int
{\cal J}[\xi]$ are invariant under diffeomorphisms and the local Lorentz
group. This construction generalizes the results known for models that are
invariant under the diffeomorphism group only \cite{Wald}, and improves
and clarifies the earlier facts about the conserved quantities associated
with a vector field
\cite{Komar1,Komar2,Trautman73,Trautman05,Benn,j1,j2,FFFR99,Fer03,A00a,A00b}
that were discovered for specific Lagrangians (usually, for the
Hilbert-Einstein one) and for specific types of vector fields (usually,
for Killing or generalized Killing ones). It is also possible to extend
the analysis to the case of quasi-invariant models that are
described by Lagrangians which change by a total derivative under the
action of the symmetry groups \cite{noninv,conserved}. In this case, however,
there seems to be no clear way to define invariant conserved currents.

Several remarks are in order. We use the physical terminology throughout 
this paper. In particular, we refer to the equation $dJ = 0$
as a conservation law. Being an $(n-1)$-form, in the local coordinates
$\{x^i\}$ the current can be expanded as $J = {\cal J}^i\epsilon_i$ 
with respect to the natural basis $\epsilon_i = {\frac 1 {(n-1)!}}\,
\epsilon_{ij_1\dots j_{n-1}}dx^{j_1}\wedge\dots\wedge dx^{j_{n-1}}$ of the 
$(n-1)$-forms (where $\epsilon_{i_1\dots i_n}$ is the totally antisymmetric 
Levi-Civita symbol on an $n$-manifold). Then $dJ = 0$ is equivalent to
the divergence equation $\partial_i{\cal J}^i = 0$ for the components
of the vector density ${\cal J}^i$. In physics, ${\cal J}^i$ is called 
a conserved current when it satisfies $\partial_i{\cal J}^i = 0$. In the
mathematical language, a form with the property $dJ = 0$ is called 
closed, but we prefer to use the standard physical terminology. 

It is well known that in diffeomorphism-invariant models one can associate 
a conserved  current to a vector field $\xi$ on the spacetime manifold.
This can be done in various ways. One example is to consider a symmetric 
energy-momentum tensor $T_j{}^i$ and a {\it Killing} vector field $\xi = \xi^i
\partial_i$ that generates an isometry of the spacetime. Since the 
energy-momentum tensor is {\it covariantly} conserved in
diffeomorphism-invariant theories, one then straightforwardly verifies that
${\cal J}^i = \xi^j T_j{}^i\sqrt{-g}$ is a conserved current, i.e.,
$\partial_i{\cal J}^i =0$. Further example is provided by a general scheme
\cite{Wald} in which a conserved
current $(n-1)$-form is derived for any solution of a diffeomorphism-invariant 
model even when $\xi$ is not a Killing field. All such $(n-1)$-form currents 
are scalars under general coordinate transformations, as
well as the corresponding charges derived from them. They proved to be useful
for the computation of the total energy and angular momentum of various
gravitational field configurations, and for the discussion of the thermodynamic
laws of the black holes in gravity models.  

The interpretation of the vector field $\xi$ is an important geometrical
and physical issue. For example, when $\xi$ is timelike, the corresponding
charge has the meaning of the energy of the gravitating system with 
respect to an observer moving along the integral lines of $\xi$, with 4-velocity
$u=\xi/|\xi|$, cf. \cite{KLB06}. In this way, the dependence of the 
charges on $\xi$ describes the usual dependence of the energy of a system on
the choice, and on the dynamics, of a physical observer. In particular, the 
invariant charge ${\cal Q}[\xi]=\int_S {\cal J}^i\epsilon_i$ is then the
integral of the {\it projection} of the energy-momentum along the vector $\xi$.
This charge reduces to the usual expression $\int_S T_0{}^0\sqrt{-g}\,
dx^1\wedge\dots\wedge dx^{n-1}$ in coordinates adapted to $\xi$ such that
$\xi=\partial_0$, and the hypersurface $S$ is defined by $x^0=$ constant. Such
an approach to the definition of the energy of a gravitating system as a 
scalar (that is, {\it invariant} depending on some vector field) closely 
follows the well-known construction for point particles (see Sect. 2.8, and in
particular Eq. (2.29) of \cite{MTW73}). 

However, the situation becomes more complicated when, besides the diffeomorphism 
symmetry, the gravitational model is also invariant under some additional gauge
group. For example, there is a large class of theories which are invariant under 
the local Lorentz group $SO(1,n-1)$, including the gauge gravity models 
\cite{Trautman05}, the supergravity, and the so called first order formulation 
of standard general relativity. The problem of defining the corresponding 
conserved quantities associated with a vector field was analyzed  previously 
\cite{Trautman05,Benn,j1,j2,FFFR99,Fer03,A00a} for specific Lagrangians (usually, 
for the Hilbert-Einstein one) and for specific types of vector fields (usually, 
for Killing or generalized Killing ones). Moreover, the resulting conserved 
quantities were often discovered to be {\it not invariant} under the local 
Lorentz group (e.g., in \cite{A00a}). 

We have shown in \cite{invar} that for models with arbitrary Lorentz-invariant 
Lagrangians it is possible to define \textit{invariant} conserved 
currents for every vector field $\xi$. These conserved currents do not 
depend on the coordinate system or the tetrad frame used to compute 
them. They depend only on the field configuration and on the choice of the 
vector field $\xi$. 

In the present paper, we further develop our approach and replace the
local Lorentz group with an arbitrary gauge Lie group $G$. This includes
two subcases: (i) the group $G$ acts in a space of internal degrees of
freedom, and (ii) the group $G$ acts on the spacetime manifold and the
local Lorentz group is its subgroup $SO(1,n-1)\in G$. The particular choice
of the general linear group $G = GL(n,R)$ is of special importance since
this group underlies the gauge formulation of the so called metric-affine
gravity (MAG) theory. As a result, we give a general construction of
the invariant conserved quantities for gravity theories with general
coordinate (diffeomorphism) and local gauge $G$ symmetries. We again
show that for every vector field $\xi$ on spacetime, and any Lagrangian
of gravitational plus matter fields there exists a conserved current
$(n-1)$-form that is a true scalar under both diffeomorphisms and $G$.

There is an important difference between the cases when $G$ is an internal
symmetry group and when it is the general linear group. In the latter
case, i.e., in the framework of MAG, there exist a well defined way to 
construct the conserved current by making use of the Yano choice of the 
generalized Lie derivative that appears in the Lagrange-Noether derivation 
of the conservation law. Such a Yano derivative (see \cite{Yano}, and cf. 
the previous derivations in \cite{invar}) is always defined in MAG in terms
of the coframe, which is a dynamical field (``translational potential")
in this approach. In contrast to this situation, in the models where $G$
is an internal symmetry group, one cannot come up with a suitable
counterpart of the Yano derivative, in general. Sometimes, a similar
construction is possible, and we explicitly demonstrate this for the case
of $G=U(1)$, for the models that include a Higgs-type complex scalar
field. However, at the moment it is unclear whether an appropriate
non-Abelian generalization can be found with the help of suitable
Higgs multiplet.

Previously \cite{invar}, we have demonstrated that our approach works nicely
for the calculation of the total mass (energy) and the total angular momentum
for the solutions of the gravitational field equations without and with
torsion. The corresponding resulting values of the mass and angular momentum
were shown to be consistent with the calculations obtained by
alternative methods, see for example \cite{Blag,Szabados} and the references 
therein. In order to test the generalized formalism, we apply it to
the analogous computation of the mass and angular momentum of the exact
solutions in MAG with nontrivial torsion and nonmetricity.

Our general notations are the same as in \cite{PR}. In
particular, we use the Latin indices $i,j,\dots$ for local
holonomic spacetime coordinates and the Greek indices $\alpha,\beta,
\dots$ label (co)frame components. Particular frame components are
denoted by hats, $\hat 0, \hat 1$, etc. As usual, the exterior
product is denoted by $\wedge$, while the interior product of a
vector $\xi$ and a $p$-form $\Psi$ is denoted by $\xi\rfloor\Psi$.
The vector basis dual to the frame 1-forms $\vartheta^\alpha$ is
denoted by $e_\alpha$ and they satisfy
$e_\alpha\rfloor\vartheta^\beta
=\delta^\beta_\alpha$. Using local coordinates $x^i$, we have
$\vartheta^\alpha=h^\alpha_idx^i$ and
$e_\alpha=h^i_\alpha\partial_i$.
We define the volume $n$-form by $\eta:=\vartheta^{\hat{0}}\wedge
\cdots\wedge\vartheta^{\hat{n}}$. Furthermore, with the help of
the interior product we define $\eta_{\alpha}:=e_\alpha
\rfloor\eta$, $\eta_{\alpha\beta}:=e_\beta\rfloor\eta_\alpha$,
$\eta_{\alpha\beta\gamma}:= e_\gamma\rfloor\eta_{\alpha\beta}$,
etc., which are bases for $(n-1)$-, $(n-2)$- and $(n-3)$-forms,
etc., respectively. Finally, $\eta_{\alpha_1\cdots \alpha_n} =
e_{\alpha_n}\rfloor\eta_{\alpha_1\cdots\alpha_{n-1}}$ is the
Levi-Civita tensor density. The $\eta$-forms satisfy the identities:
\begin{eqnarray}
\vartheta^\beta\wedge\eta_\alpha &=& \delta_\alpha^\beta\eta ,\\
\vartheta^\beta\wedge\eta_{\mu\nu} &=& \delta^\beta_\nu\eta_{\mu} -
\delta^\beta_\mu\eta_{\nu},\label{veta1}\\ \label{veta}
\vartheta^\beta\wedge\eta_{\alpha\mu\nu}&=&\delta^\beta_\alpha
\eta_{\mu\nu} + \delta^\beta_\mu\eta_{\nu\alpha} +
\delta^\beta_\nu\eta_{\alpha\mu},\\
\vartheta^\beta\wedge\eta_{\alpha\gamma\mu\nu}&=&\delta^\beta_\nu
\eta_{\alpha\gamma\mu} - \delta^\beta_\mu\eta_{\alpha\gamma\nu}
+ \delta^\beta_\gamma\eta_{\alpha\mu\nu} -
\delta^\beta_\alpha\eta_{\gamma\mu\nu},\label{veta2}
\end{eqnarray}
etc. The line element $ds^2 = g_{\alpha\beta}\,\vartheta^\alpha\otimes
\vartheta^\beta$ is defined by the spacetime metric $g_{\alpha\beta}$
of signature $(+,-,\cdots,-)$.

\section{General formalism in condensed notation}

Let us consider the most general case of a gravitational theory with the
diffeomorphism and an additional arbitrary gauge symmetry. As a matter
of fact, all gravitational theories are special cases of metric-affine
gravity (MAG) with the gravitational field described by the three basic
variables: the metric $g_{\alpha\beta}$,
the coframe $\vartheta^\alpha$, and the linear connection $\Gamma_\alpha
{}^\beta$. In addition, the nongravitational sector of the theory
contains the usual matter fields $\psi$ (scalars and spinors of
any rank that describe massive particles) and the gauge fields $A$
of a certain internal symmetry group (these are 1-forms with values
in the corresponding Lie algebra, or $p$-forms, in general).

Taking this into account, we denote the total symmetry group by $G$. It is
defined as a direct product of the {\it internal symmetry} group and the
{\it spacetime symmetry} group. The former can be any Abelian or non-Abelian
Lie group, acting via a suitable representation on the multiplet of
matter fields. The spacetime symmetry group can be a Lorentz, linear,
conformal, de Sitter group or other, acting on the geometric objects on
the spacetime manifold. Accordingly, we will denote the gauge fields of
internal symmetries $A$ together with the gravitational connection
$\Gamma$ by a collective gauge field ${\cal A}^a$ which is a 1-form taking
values in the Lie algebra ${\cal G}$ of the group $G$. The index $a$ runs
through both the gravitational and nongravitational sectors and labels the
corresponding  infinitesimal parameters $\varepsilon^a$  of ${\cal G}$.
In a similar way, we will denote all the covariant fields of the model
by a collective field $\Psi$. The latter includes the gravitational sector
(for example, $g_{\alpha\beta}, \vartheta^\alpha$) and the matter fields
$\psi^A$. Finally, we collect {\it all} the fields of the model
into a single object $\Phi^I = \{\Psi,{\cal A}\}$, where the
index $I$ runs over all the components of the fields in all sectors.

After these preliminaries, we can describe the generalization of the
construction \cite{invar} of conserved currents for theories invariant
under diffeomorphisms and an arbitrary gauge group $G$. As before, we start
with the total Lagrangian $n$-form $V^{\rm tot}(\Phi^I, d\Phi^I)$ and define 
the generalized field momenta and the energy terms by
\begin{equation}
H_I := -\,{\frac {\partial V^{\rm tot}}{\partial d\Phi^I}},\qquad
E_I := {\frac {\partial V^{\rm tot}}{\partial \Phi^I}}.
\end{equation}
The total variation of the Lagrangian then reads
\begin{equation}
\delta V^{\rm tot} = \delta\Phi^I\wedge {\cal F}_I -
d(\delta\Phi^I\wedge H_I),\label{deltaV3}
\end{equation}
where the variational derivative is defined by
\begin{equation}
{\cal F}_I := {\frac {\delta V^{\rm tot}}{\delta \Phi^I}} =
(-1)^{p(I)}DH_I + E_I.
\end{equation}
Here $p(I)$ denotes the rank (in the exterior sense) of the
corresponding sector of the collective field.

The equation (\ref{deltaV3}) describes how the Lagrangian changes under 
the change of the fields. This includes two cases: (i) when the variations
of the fields are arbitrary, and (ii) when the fields are transformed under the
action of the symmetry group. In the first case, when the action is stationary 
(i.e., $\delta V^{\rm tot} = 0$) under the {\it arbitrary} variations
$\delta\Phi^I$, one finds from (\ref{deltaV3}) the field equations
\begin{equation}
{\cal F}_I = 0.\label{onshell}
\end{equation}

However, in the second case the variation $\delta\Phi^I$ is not arbitrary, and
the invariance of the action gives rise to the conservation laws. The total 
variation of the field variables under diffeomorphisms and the gauge symmetry reads
\begin{equation}
\delta\Phi^I = \ell_{(\varsigma\xi)}\Phi^I + \delta_{(\varsigma\varepsilon)}
\Phi^I =: \varsigma\,{\cal L}_{\xi,\varepsilon}\Phi^I.\label{LieD}
\end{equation}
Here $\varsigma$ is an arbitrary infinitesimal constant parameter, $\ell_\xi
= \xi\rfloor d + d \xi\rfloor$ is the ordinary Lie derivative, and
the second term describes the gauge transformation
\begin{equation}
\delta_{\varsigma\varepsilon}\Phi^I = \varsigma\left[\varepsilon^a
(\rho_a)^I{}_J\,\Phi^J - (\sigma_a)^I d\varepsilon^a\right].\label{gauge}
\end{equation}
This reduces to the well known law for the case of the Lorentz symmetry.
The generators $\rho_a$ satisfy the commutation relation $[\rho_a,\rho_b]^I
{}_J = f^c{}_{ab}\,(\rho_c)^I{}_J$ with the structure constants $f^c{}_{ab}$
of the Lie algebra ${\cal G}$.
It is worthwhile to note that $(\sigma_a)^I = \delta^I_a$ is a unit
matrix in the gauge sector and is trivial in the covariant sector.

We will call ${\cal L}_{\xi,\varepsilon}\Phi^I$ defined by (\ref{LieD})
a generalized Lie derivative of the multifield $\Phi^I$.

After these preliminary steps, we can derive the conservation law and
introduce the conserved current. Namely, for the general models under
consideration, we define the generalized current
\begin{equation}
J[\xi,\varepsilon] := \xi\rfloor V^{\rm tot} +
({\cal L}_{\xi,\varepsilon}\Phi^I)\wedge H_I.\label{JdefG}
\end{equation}
As for the case of the local Lorentz symmetry (cf. \cite{invar}), it
satisfies
\begin{equation}
dJ[\xi,\varepsilon] \equiv ({\cal L}_{\xi,\varepsilon}\Phi^I)
\wedge{\cal F}_I.
\end{equation}
This is just the total variation formula (\ref{deltaV3}) in a different
form. By using the Noether identities (which we do not write down here
explicitly, see the subsequent discussion of the internal symmetry in
Sec.~\ref{internal} and of the general linear group in Sec.~\ref{external}),
we can recast this current as
\begin{equation}
J[\xi,\varepsilon] \equiv d\left(\Xi^I[\xi,\varepsilon]\wedge H_I\right)
+ \Xi^I[\xi,\varepsilon]\wedge {\cal F}_I.\label{JdG}
\end{equation}
Here we denoted 
\begin{equation}
\Xi^I[\xi,\varepsilon] := \xi\rfloor\Phi^I - \varepsilon^a
\delta_a^I.\label{XiG}
\end{equation}

When the field equations (\ref{onshell}) are satisfied, the generalized
current (\ref{JdefG}) is conserved, $dJ[\xi,\varepsilon] = 0$, and hence
one can define the corresponding charge by the integrals
\begin{equation}
Q[\xi,\varepsilon] = \int_S J[\xi,\varepsilon] =
\int_{\partial S} \Xi^I[\xi,\varepsilon]\wedge H_I\label{chargeG}
\end{equation}
over a spacelike $(n-1)$-hypersurface $S$ with an $(n-2)$-dimensional boundary
$\partial S$. The equation (\ref{chargeG}) arises, as usual, when we integrate 
the conservation law $d J[\xi,\varepsilon] = 0$ over the $n$-volume domain 
with the boundary $S_1 + S_2 + T$, where $S_1$ and $S_2$ are $(n-1)$-dimensional
spacelike hypersurfaces (which correspond to the arbitrary time values $t_1$ 
and $t_2$, respectively) and $T$ is a timelike surface that connects them. 
Then assuming that the fields satisfy the boundary conditions such that 
$\int_T J = 0$, we find that $Q[\xi,\varepsilon] = \int_{S(t)} J[\xi,
\varepsilon]$ is constant. Using subsequently (\ref{JdG}), one finds 
(\ref{chargeG}) with the help of the Stokes theorem. We will always assume
the appropriate asymptotic behavior of the geometric and matter fields 
that makes the conserved charges well-defined objects. Since we deal with
a general scheme without specifying a Lagrangian, the necessary boundary 
conditions appropriate for all theories cannot be given explicitly. They
depend on the particular model and should be chosen after a case by case 
inspection of a theory. 

The functions $\varepsilon^a$ parametrize a {\it family} of conserved
currents (\ref{JdefG}) and charges (\ref{chargeG}) associated with a vector
field $\xi$. In order to select {\it invariant} conserved quantities, we
will have to specialize to a particular choice of $\varepsilon$. The trivial
choice $\varepsilon^a=0$ yields a {\it noninvariant} current and charge.
Indeed, then ${\cal L}_{\{\xi,\varepsilon\}}\Psi = \ell_\xi\Psi$, and
the last term in (\ref{JdefG}) is not gauge invariant, since the ordinary
Lie derivative $\ell_\xi$ of a covariant object is not covariant under
gauge transformations. In \cite{invar} it was shown that one can define
covariant conserved currents (and charges) with the help of an appropriate
choice of $\varepsilon(\xi)$. Unfortunately, there seems to be no
general recipe how to choose $\varepsilon(\xi)$ for an arbitrary internal
and external symmetry group. Accordingly, the situation should be studied
on a case by case basis. In the next sections we analyze separately
the models with internal gauge symmetries (Sec.~\ref{internal}) and the
metric-affine gravity (Sec.~\ref{external}).

\section{Conserved currents in gauge theories}\label{internal}

First we consider the class of theories that are invariant under the
diffeomorphism group and an arbitrary local (gauge) Lie group $G$. We
denote the corresponding Lie algebra $\cal G$. Its generators $\rho_a$,
$a=1,\dots,{\rm dim}({\cal G})$, satisfy the commutator relations
\begin{equation}
 \left[\rho_a,\rho_b\right] = f^c{}_{ab}\,\rho_c,
\end{equation}
with the structure constants $f^c{}_{ab}$. We also consider matter $p$-form
fields $\Psi^A$ transforming covariantly under the action of $G$, such that
\begin{equation}
\delta_{\varepsilon}\Psi^A = -\,\varepsilon^a (\rho_a)^A{}_B\Psi^B,
\end{equation}
where $\varepsilon^a(x)$ are the parameters of the transformation,
and $(\rho_a)^A{}_B$ denote the matrix representation of the generators $\rho_a$
acting in the vector space of the matter fields $\Psi^A$. The covariant 
derivative is then defined by
\begin{equation}
 D\Psi^A := d\Psi^A - A^a(\rho_a)^A{}_B\wedge\Psi^B,
\end{equation}
where $A^a$ denotes the gauge field potential 1-form. The corresponding
gauge field strength reads
\begin{equation}
F^a := dA^a-\frac{1}{2}f^a{}_{bc}\,A^b\wedge A^c.
\end{equation}
The latter, as usual, may be derived from the commutator of the covariant
derivatives, $DD\Psi^A = -\,F^a(\rho_a)^A{}_B\Psi^B$.
For completeness, let us recall the transformation laws of the potential
and the field strength:
\begin{eqnarray}
\delta_{\varepsilon}A^a&=& - d\varepsilon^a+f^a{}_{bc}
A^b\varepsilon^c,\\ 
\delta_{\varepsilon} F^a &=& - \varepsilon^b f^a{}_{bc}F^c.
\end{eqnarray}

Now, let $V(\Psi^A,A^a,D\Psi^A,F^a)$ be a general Lagrangian $n$-form.
We define
\begin{equation}
H_A := -\,{\frac {\partial V}{\partial D\Psi^A}},\qquad
E_A := {\frac {\partial V}{\partial \Psi^A}},\label{HE}
\end{equation}
\begin{equation}
H_a := -\,\frac {\partial V}{\partial F^a},\qquad
E_a := \frac{\partial V}{\partial A^a}.\label{HE2}
\end{equation}
Then a total variation of the Lagrangian is
\begin{equation}
\delta V = \delta\Psi^A\wedge {\cal F}_A +\delta A^a\wedge {\cal F}_a
- d(\delta\Psi^A\wedge H_A+\delta\Psi^a\wedge H_a),\label{deltaV1}
\end{equation}
where we introduced the variational derivatives
\begin{equation}
{\cal F}_A := {\frac {\delta V}{\delta \Psi^A}} = (-1)^{p}DH_A + E_A.
\qquad {\cal F}_a := \frac{\delta V}{\delta A^a} = -DH_a + E_a.
\end{equation}
Here $p$ denotes the rank (in the exterior sense) of the matter field
$\Psi^A$.
Assuming that the action is stationary for the arbitrary variations of the
fields, from the above we derive the system of field equations
\begin{equation}
{\cal F}_A =0,\qquad {\cal F}_a =0.\label{onshell1}
\end{equation}

We assume that the action of the theory is invariant under diffeomorphism and
gauge transformations. The total {\it infinitesimal symmetry variation} of
the
dynamical fields then consists of two terms:
\begin{eqnarray}
\delta\Psi^A &=& \varsigma{\cal L}_{\{\xi,\varepsilon\}}\Psi^A:=
\ell_{(\varsigma\xi)}\Psi^A +\delta_{(\varsigma\varepsilon)}
\Psi^A , \label{defL}\\
\delta A^a &=& \varsigma{\cal L}_{\{\xi,\varepsilon\}} A^a:=
\ell_{(\varsigma\xi)}A^a +\delta_{(\varsigma\varepsilon)} A^a .
\end{eqnarray}
The first terms on the right-hand sides come from a
diffeomorphism generated by a vector field $\xi$, and $\ell_\xi$ is
the Lie derivative along that field.

Putting $\xi =0$ and assuming $\varepsilon^a$ completely arbitrary, we
straightforwardly derive from (\ref{deltaV1}) the Noether identities
for the gauge symmetry:
\begin{eqnarray}
E_a\equiv (\rho_a)^A{}_B\Psi^B\wedge H_B-f^b{}_{ca}A^c\wedge H_b,\\
dE_a - (\rho_a)^A{}_B\Psi^B\wedge {\cal F}_A + f^b{}_{ca}A^c
\wedge{\cal F}_b\equiv 0.
\end{eqnarray}
Analogously, putting $\varepsilon^a =0$ and assuming an arbitrary vector
field $\xi$, we find from (\ref{deltaV1}) the Noether identities for
the diffeomorphism symmetry:
\begin{eqnarray}
e_\alpha\rfloor V+e_\alpha\rfloor d\Psi^A\wedge H_A+e_\alpha\rfloor
dA^a\wedge H_a \equiv e_\alpha\rfloor\Psi^A\wedge E_A+e_\alpha\rfloor
A^a\wedge E_a,\label{Ndiff1}\\
e_\alpha\rfloor d\Psi^A\wedge{\cal F}_A+(-1)^p e_\alpha\rfloor\Psi^A
\wedge d{\cal F}_A+e_\alpha\rfloor dA^a\wedge{\cal F}_a
-e_\alpha\rfloor A^a\wedge d{\cal F}_a\equiv 0.\label{Ndiff2}
\end{eqnarray}
The last identity is equivalent to
\begin{equation}
\ell_{e_\alpha}\Psi^A\wedge{\cal F}_A+\ell_{e_\alpha}A^a\wedge{\cal F}_a
\equiv d\left(e_\alpha\rfloor\Psi^A\wedge{\cal F}_A+e_\alpha\rfloor A^a
\wedge{\cal F}_a\right).\label{Ndiff3}
\end{equation}

After these preliminaries, we are now in a position to derive the
generalised conserved current associated with any vector field $\xi$.
The condition of the invariance of the theory under a general variation
(\ref{defL}) follows directly from (\ref{deltaV1}) and reads
\begin{eqnarray}
d(\xi\rfloor V) &=& ({\cal L}_{\{\xi,\varepsilon\}}\Psi^A)\wedge{\cal F}_A
+({\cal L}_{\{\xi,\varepsilon\}}A^a)\wedge{\cal F}_a\nonumber\\
&& -\,d\left[({\cal L}_{\{\xi,\varepsilon\}}\Psi^A)\wedge H_A+({\cal
L}_{\{\xi,\varepsilon\}}A^a)\wedge H_a\right].\label{deltaV2}
\end{eqnarray}
Introducing the current $(n-1)$-form
\begin{equation}
J[\xi,\varepsilon] := \xi\rfloor V +
({\cal L}_{\{\xi,\varepsilon\}}\Psi^A)\wedge H_A
+({\cal L}_{\{\xi,\varepsilon\}}A^a)\wedge H_a,\label{Jdef}
\end{equation}
we see from (\ref{deltaV2}) that
\begin{equation}
dJ[\xi,\varepsilon] = ({\cal L}_{\{\xi,\varepsilon\}}\Psi^A)
\wedge{\cal F}_A+({\cal L}_{\{\xi,\varepsilon\}}A^a)
\wedge{\cal F}_a.
\end{equation}
Hence, this current is conserved, $dJ[\xi,\varepsilon] = 0$, for any $\xi$
and $\varepsilon^a$, when the field equations (\ref{onshell1}) are satisfied.

Using (\ref{Jdef}), (\ref{defL}) and the Noether identities of the
diffeomorphism symmetry (\ref{Ndiff1})-(\ref{Ndiff3}),
we rewrite the current (\ref{Jdef}) as
\begin{equation}
J[\xi,\varepsilon] = d\Pi[\xi,\varepsilon] + \xi\rfloor\Psi^A\wedge
{\cal F}_A+ \Xi^a[\xi,\varepsilon]\wedge {\cal F}_a.\label{Jd}
\end{equation}
Here we introduced the superpotential $(n-2)$-form
\begin{equation}
\Pi[\xi,\varepsilon] := \xi\rfloor\Psi^A\wedge
H_A+\Xi^a[\xi,\varepsilon]\wedge H_a \label{Pi}
\end{equation}
and defined [cf. the general definition (\ref{XiG})]
\begin{equation}
\Xi^a[\xi,\varepsilon] := \xi\rfloor A^a - \varepsilon^a.\label{Xi}
\end{equation}
Accordingly, on the solutions of the field equations (\ref{onshell1}),
the conserved charge is computed as an integral over an $(n-2)$-boundary:
\begin{equation}
Q[\xi,\varepsilon] = \int_S J[\xi,\varepsilon] =
\int_{\partial S} \Pi[\xi,\varepsilon].\label{charge}
\end{equation}
As before, the functions $\varepsilon^a$ parametrize a family of conserved
currents and charges associated with a vector field $\xi$. These conserved
quantities are not scalars, in general. They are invariant under the
diffeomorphisms, but they become invariant under local gauge transformations
only for certain special choices of the parameters $\varepsilon^a(\xi)$.
One choice that is always possible is to take a {\it nondynamical} (or
background) gauge field $\overline{A}^a$ and define $\varepsilon^a =
\xi\rfloor\overline{A}^a$. Then $\Xi^a = \xi\rfloor(A^a - \overline{A}^a)$
is obviously a gauge-covariant quantity, and consequently the conserved
current and charge are true scalars.

Other choices are also possible, in general, with a covariant $\Xi^a$
constructed from the dynamical fields available in a particular model.
The simplest is a ``natural" choice with $\varepsilon^a = \xi\rfloor A^a$
but then $\Xi^a = 0$, and the corresponding contribution to the current
is trivial. As another example, recall \cite{invar} that we have
demonstrated how one can use the coframe field in order to define the
so-called Yano derivative which leads to the invariant conserved currents.
A similar construction is outlined in the next section for the Abelian
gauge field model.

\section{Abelian model: defining a covariant $\Xi$}

Let us consider an Abelian model with the local gauge symmetry group
$G=U(1)$. Then, besides the choice $\varepsilon=\xi\rfloor
\overline{A}$, a nontrivial field $\varepsilon(\xi)$ can be defined,
provided a $U(1)$-covariant scalar field $\phi$ is available. We assume
that under a $U(1)$ transformation with parameter $\lambda$ (with $\lambda
\in [0,2\pi]$), the field $\phi$ transforms as $\phi'=e^{i\lambda}\phi$.
Using $\phi$, we can construct $\varepsilon(\xi)$ from the assumption that
\begin{equation}
{\cal L}_{\xi,\varepsilon}\phi=0. \label{Lphi0}
\end{equation}
A 0-form $\phi$ (a complex scalar field) usually plays the role of a Higgs
field in such models. The condition (\ref{Lphi0}) reads explicitly
\begin{equation}
 \ell_\xi\phi+i\varepsilon\phi=0 \label{lphiephi}
\end{equation}
By multiplying this by $\phi^\dagger$, we solve (\ref{lphiephi}) for
$\varepsilon$. This yields
\begin{equation}
 \varepsilon=i\frac{\phi^\dagger\ell_\xi\phi}{\phi^\dagger\phi}. \label{epu1}
\end{equation}
The solution $\varepsilon$ is real, for every $\xi$, when  $\phi^\dagger
\phi = $const. Then we can always choose the normalization such that
$\phi^\dagger\phi=1$, and recast (\ref{epu1}) as
\begin{equation}
 \varepsilon=\frac{i}{2}
\left[\phi^\dagger(\ell_\xi\phi)-(\ell_\xi\phi^\dagger)\phi\right]=\frac{i}{2}
\xi\rfloor\left[\phi^\dagger(d\phi)-(d\phi^\dagger)\phi\right].
\label{epu12}
\end{equation}
It is interesting to notice that the right-hand side of (\ref{epu12})
is proportional to the $U(1)$ current of a free complex scalar field $\phi$.

\subsection{Complex scalar field on a background spacetime}

Consider now the model with a complex scalar field $\psi$ coupled to the
electromagnetic field $A$ and the gravitational field. The latter can be
though treated as a curved background, since at the moment we are concerned
primarily with the aspects of the local $U(1)$ invariance. The covariant
derivative is given, as usual, by $D\psi = d\psi+iA\psi$. Let us take the
total Lagrangian $V^{\rm tot}=V^{\rm tot}(\psi,D\psi,dA)$, where we have
written
explicitly only the dynamical fields $\psi,A$, whereas a fixed metric $g$
is assumed.
Since $\psi$ is a 0-form, the superpotential (\ref{Pi}) reduces to
\begin{equation}
\Pi[\xi,\varepsilon] = \left(\xi\rfloor A -\varepsilon\right) H,
\qquad H = -\frac{\partial V}{\partial F}.
\end{equation}
Clearly, in this case the ``natural'' choice $\varepsilon=\xi\rfloor A$ leads
to trivial conserved quantities.

However, we can use the dynamical (Higgs type) field $\psi$ to construct a
nontrivial conserved quantity using (\ref{epu12}). If we take
\begin{equation}
\phi=\frac{\psi}{\sqrt{\psi^\dagger\psi}},
\end{equation}
then $\phi^\dagger\phi=1$ and we obtain
\begin{equation}
 \varepsilon=\frac{i}{2}\,
\xi\rfloor\left[\frac{\psi^\dagger(d\psi)-(d\psi^\dagger)\psi}{\psi^\dagger
\psi}\right]=\frac{m}{e}\frac{1}{\psi^\dagger\psi}\,\xi\rfloor j_{\rm free},
\label{epu13}
\end{equation}
where
$j_{\rm free} = \frac{ie}{2m}\left[\psi^\dagger(d\psi)-(d\psi^\dagger)
\psi\right]$ is the electric current density 1-form of a free complex
scalar field ($\hbar=c=1$). Using (\ref{epu13}), we find
\begin{equation}
\Xi = \xi\rfloor A-\varepsilon=-\frac{m}{e}\frac{1}{\psi^\dagger\psi}
\,\xi\rfloor j,
\end{equation}
where now
$j = \frac{ie}{2m}\left[\psi^\dagger(D\psi)-(D\psi^\dagger)\psi\right]$ is
the invariant current of the field $\psi$ interacting with the
electromagnetic
field. Thus, we obtain finally
\begin{equation}
\Pi[\xi] = -\,\frac{m}{e}\,\frac{1}{\psi^\dagger\psi}\,(\xi\rfloor j)\,H .
\end{equation}
The corresponding conserved quantity $Q[\xi]=\int_{\partial S}\Pi[\xi]$ is
gauge invariant and a scalar under general coordinate transformations.

\section{Invariant conserved currents for metric-affine gravity}
\label{external}

The geometry of MAG (`metric-affine gravity') is described by the {\it
  curvature} two-form $R_{\alpha}{}^{\beta}$, the {\it nonmetricity}
one-form $Q_{\alpha\beta}:=-Dg_{\alpha\beta}$, and the {\it torsion}
two-form $T^{\alpha} :=D\vartheta^{\alpha}$ which are the
gravitational field strengths for the linear connection
$\Gamma_{\alpha}{}^{\beta}$, metric $g_{\alpha\beta}$, and coframe
$\vartheta^{\alpha}$, respectively. The corresponding physical sources
are the 3-forms of canonical energy-momentum $\Sigma_{\alpha}$ and
hypermomentum $\Delta^{\alpha}{}_{\beta}$.  The latter includes the
dilation, shear, and spin currents associated to matter.  The field
equations and the formalism are comprehensively described in \cite{PR,maggron}.

The MAG theory is invariant under diffeomorphisms and the local general
linear group $G = GL(n,R)$ that acts on the geometric objects (that
describe the gravitational field) as
\begin{equation}\label{deltaVGg}
\delta\vartheta^\alpha = \varepsilon^\alpha{}_\beta\,\vartheta^\beta,
\quad \delta\Gamma_\beta{}^\alpha = - D\varepsilon^\alpha{}_\beta,\quad
\delta g_{\alpha\beta} = -\,\varepsilon^\gamma{}_\alpha\,g_{\gamma\beta}
- \varepsilon^\gamma{}_\beta\,g_{\alpha\gamma},
\end{equation}
and on the matter fields $\psi^A$ as
\begin{equation}
\delta\psi^A = \varepsilon^\alpha{}_\beta\,(\rho^\beta{}_\alpha)^A_B
\,\psi^B.\label{deltapsi}
\end{equation}
Here the elements of the matrix $\varepsilon^\alpha{}_\beta(x)$
are the $n^2$ arbitrary local parameters, and $\rho^\beta{}_\alpha$ are
the generators of the general linear group in a corresponding representation.

\subsection{Lagrangian formalism}

We assume that the total Lagrangian $n$-form $V^{\rm tot}(g_{\alpha\beta},
dg_{\alpha\beta},\vartheta^{\alpha},d\vartheta^{\alpha},\Gamma_{\alpha}
{}^{\beta},d\Gamma_{\alpha}{}^{\beta},\psi^A,d\psi^A)$ is \textit{invariant}
under local transformations (\ref{deltaVGg}). Then one can verify \cite{PR}
that it always has the form
\begin{equation}
V^{\rm tot} = V^{\rm tot}(g_{\alpha\beta},Q_{\alpha\beta},\vartheta^{\alpha},
T^{\alpha},R_{\alpha}{}^{\beta},\psi,D\psi),\label{lagrVtot}
\end{equation}
where $D\psi^A$ denotes the covariant exterior derivative of the matter
field $\psi^A$. In accordance with the general scheme, we denote:
\begin{equation}
{\cal H}_{\alpha} := -\,{\frac{\partial V^{\rm tot}}{\partial T^{\alpha}}}\,,
\qquad {\cal H}^{\alpha}{}_{\beta} := -\,{\frac{\partial V^{\rm tot}}
{\partial R_{\alpha}{}^{\beta}}}\,,\qquad {\cal M}^{\alpha\beta} :=
-\,2{\frac{\partial V^{\rm tot}}{\partial Q_{\alpha\beta}}}\,.\label{HHM}
\end{equation}
Furthermore, we also introduce
\begin{equation}
{\cal E}_{\alpha} := {\frac{\partial V^{\rm
tot}}{\partial\vartheta^{\alpha}}},
\qquad {\cal E}^\alpha{}_\beta:= {\frac{\partial V^{\rm tot}}{\partial
\Gamma_{\alpha}{}^{\beta}}},
\qquad \mu^{\alpha\beta} := 2{\frac{\partial
V^{\rm tot}}{\partial g_{\alpha\beta}}}\,.\label{EEm}
\end{equation}

Then a general variation of the total Lagrangian reads
\begin{eqnarray}
\delta V^{\rm tot} =\delta\vartheta^{\alpha}\wedge {\cal F}_\alpha
+ \delta\Gamma_\alpha{}^{\beta}\wedge{\cal F}^\alpha{}_\beta +
{\frac 12}\,\delta g_{\alpha\beta}\,f^{\alpha\beta} + \delta\psi^A\wedge
{\cal F}_A\nonumber\\
+ d\left(- \delta\vartheta^{\alpha}\wedge {\cal H}_\alpha -\delta
\Gamma_{\alpha}{}^{\beta}\wedge{\cal H}^\alpha{}_\beta + {\frac 12}
\,\delta g_{\alpha\beta}{\cal M}^{\alpha\beta} + \delta\psi^A\wedge
{\frac {\partial V^{\rm tot}}{\partial D\psi^{A}}}\right), \label{var01}
\end{eqnarray}
where we have defined the variational derivatives with respect to (w.r.t.)
the
gravitational potentials:
\begin{eqnarray}
{\cal F}_\alpha &:=& {\frac{\delta V^{\rm tot}}{\delta\vartheta^{\alpha}}}
= -\,D{\cal H}_\alpha + {\cal E}_\alpha,\label{first}\\
{\cal F}^\alpha{}_\beta &:=& {\frac{\delta V^{\rm tot}}{\delta\Gamma_\alpha
{}^\beta}} = -\,D{\cal H}^\alpha{}_\beta + {\cal E}^\alpha{}_\beta,
\label{second}\\
f^{\alpha\beta} &:=& 2{\frac{\delta V^{\rm tot}}{\delta g_{\alpha\beta}}}
= -\,D{\cal M}^{\alpha\beta} + \mu^{\alpha\beta}, \label{fab}\\
{\cal F}_A &:=& {\frac{\delta V^{\rm tot}}{\delta\psi^{A}}} =
{\frac{\partial V^{\rm tot}}{\partial\psi^A}} - (-1)^p{\frac{\partial
V^{\rm tot}}{\partial D\psi^{A}}}. \label{FcA}
\end{eqnarray}

When the action is demanded to be stationary w.r.t. arbitrary variations
of the
variables, we find from (\ref{var01}) the system of {\it field equations}:
\begin{equation}
{\cal F}_\alpha =0,\qquad {\cal F}^\alpha{}_\beta =0,\qquad f^{\alpha\beta}
=0,\qquad {\cal F}_A =0.\label{onshellMAG}
\end{equation}

\subsection{Noether identities for the local $GL(n,R)$ and diffeomorphisms}

Substituting (\ref{deltaVGg}) and (\ref{deltapsi}) into (\ref{var01}), one
derives the {\it Noether identities for the local linear symmetry}:
\begin{eqnarray}\label{Nlin1}
\vartheta^\alpha\wedge{\cal F}_\beta + D{\cal F}^\alpha{}_\beta - f^\alpha
{}_\beta + (\rho^\alpha{}_\beta)^A_B\,\psi^B\wedge{\cal F}_A &\equiv&0,\\
{\cal E}^\alpha{}_\beta + \vartheta^\alpha\wedge{\cal H}_\beta +
{\cal M}^\alpha{}_\beta - (\rho^\alpha{}_\beta)^A_B\,\psi^B\wedge
{\frac{\partial V^{\rm tot}}{\partial D\psi^{A}}} &\equiv&0.\label{Nlin2}
\end{eqnarray}
The second identity can be used as a tool for the practical calculation
of ${\cal E}^\alpha{}_\beta$ (in order to circumvent a difficult direct
computation). On the other hand, the identity (\ref{Nlin1}) shows that
the so-called 0th field equation of MAG, eq. (\ref{fab}), is a consequence of
the 1st and the 2nd field equations, (\ref{first}) and (\ref{second}) and
of the
equation of motion of matter (\ref{FcA}).

For a {\it diffeomorphism} generated by an arbitrary vector field $\xi$,
we find:
\begin{eqnarray}
\ell_\xi V^{\rm tot} &=& {\frac 12}\,(\ell_{\xi}g_{\alpha\beta})
\,\mu^{\alpha\beta} - {\frac 12}\,(\ell_{\xi} Q_{\alpha\beta})\wedge
{\cal M}^{\alpha\beta}\nonumber\\
&& +\,(\ell_{\xi}\vartheta^\alpha)\wedge{\cal E}_\alpha - (\ell_\xi T^\alpha)
\wedge{\cal H}_\alpha - (\ell_\xi R_\alpha{}^\beta)\wedge
{\cal H}^{\alpha}{}_{\beta}\nonumber\\
&& +\,(\ell_\xi\psi^A)\wedge{\frac{\partial V^{\rm tot}} {\partial\psi^A}}
+ (\ell_{\xi}D\psi^A)\wedge {\frac{\partial V^{\rm tot}} {\partial
D\psi^A}}\,.
\label{vard2}
\end{eqnarray}
This equation does not look invariant under the action of the local
$GL(n,R)$.
However, if we make an actual linear transformation of all the fields in
(\ref{vard2}) by substituting (\ref{deltaVGg}) and (\ref{deltapsi}), we find
that the right-hand side is changed by
\begin{eqnarray}
(\xi\rfloor d\varepsilon^\beta{}_\alpha)\Bigg[\vartheta^\alpha\wedge
{\cal F}_\beta + D{\cal F}^\alpha{}_\beta - f^\alpha{}_\beta
+ (\rho^\alpha{}_\beta)^A_B\,\psi^B\wedge{\cal F}_A \nonumber\\
- D\left({\cal E}^\alpha{}_\beta + \vartheta^\alpha\wedge{\cal H}_\beta
+ {\cal M}^\alpha{}_\beta - (\rho^\alpha{}_\beta)^A_B\,\psi^B\wedge
{\frac{\partial V^{\rm tot}}{\partial D\psi^{A}}}\right)\Bigg].
\end{eqnarray}
This is zero in view of the Noether identities (\ref{Nlin1}) and
(\ref{Nlin2}).

\subsection{Generalized Lie derivatives and covariant Noether identities}

As a result, we can proceed like in the previous paper \cite{invar}, and
take an \textit{arbitrary $gl(n,R)$-valued 0-form} $B_\alpha{}^\beta
(\xi)$ and add to (\ref{vard2}) a {\it zero term},
\begin{eqnarray}
B_\alpha{}^\beta\Bigg[\vartheta^\alpha\wedge
{\cal F}_\beta + D{\cal F}^\alpha{}_\beta - f^\alpha{}_\beta
+ (\rho^\alpha{}_\beta)^A_B\,\psi^B\wedge{\cal F}_A \nonumber\\
- D\left({\cal E}^\alpha{}_\beta + \vartheta^\alpha\wedge{\cal H}_\beta
+ {\cal M}^\alpha{}_\beta - (\rho^\alpha{}_\beta)^A_B\,\psi^B\wedge
{\frac{\partial V^{\rm tot}}{\partial D\psi^{A}}}\right)\Bigg].\label{zero}
\end{eqnarray}
As it is easily verified, this addition is {\it equivalent} to the
replacement
of the usual Lie derivative $\ell_\xi$ with a {\it generalized Lie
derivative}
$L_\xi:=\ell_\xi + B_\beta{}^\alpha\rho^\beta{}_\alpha$ when applied to all
geometric an matter fields. This generalized derivative will be
\textit{covariant} provided $B_\alpha{}^\beta$ transforms according to
\begin{equation}
B'_\alpha{}^\beta=(L^{-1})^\rho{}_\alpha B_\rho{}^\gamma
L^\beta{}_\gamma -(L^{-1})^\gamma{}_\alpha (\xi\rfloor
dL^\beta{}_\gamma),\label{TB}
\end{equation}
when the coframe is changed as $\vartheta^\alpha\rightarrow\vartheta'^\alpha
= L^\alpha{}_\beta \vartheta^\beta$, with $L^\alpha{}_\beta\in
GL(n,R)$.

There are three convenient choices: (i) $B_\beta{}^\alpha = \xi\rfloor
\Gamma_\beta{}^\alpha$ with the dynamical linear connection $\Gamma_\beta
{}^\alpha$, (ii) $B_\beta{}^\alpha = \xi\rfloor{\stackrel{\{\,\}}
{\Gamma}}_\beta{}^\alpha$ with the Riemannian connection ${\stackrel{\{\,\}}
{\Gamma}}_\beta{}^\alpha$, and (iii) the Yano choice  $B_\beta{}^\alpha =
- e_\beta\rfloor\ell_\xi\vartheta^\alpha$. Each of these options give rise to
the covariant Lie derivatives:
\begin{eqnarray}
{\hbox{\L}}_\xi &:=&\ell_\xi + \xi\rfloor\Gamma_\beta{}^\alpha
\rho^\beta{}_\alpha=\xi\rfloor D + D\xi\rfloor,\label{lie1}\\
{\stackrel{\{\,\}}{\hbox{\L}}}_\xi &:=&\ell_\xi +
\xi\rfloor{\stackrel{\{\,\}}
{\Gamma}}_\beta{}^\alpha\rho^\beta{}_\alpha =
\xi\rfloor{\stackrel{\{\,\}}{D}}
+ {\stackrel{\{\,\}}{D}}\xi\rfloor,\label{lie2}\\
{\cal L}_\xi &:=& \ell_\xi - \Theta_\beta{}^\alpha
\rho^\beta{}_\alpha,\quad {\rm with}\quad \Theta_\alpha{}^\beta
:= e_\alpha\rfloor\ell_\xi\vartheta^\beta.\label{lie3}
\end{eqnarray}

We are free to use in (\ref{vard2}) any generalised (covariant) Lie
derivative
instead of the usual (noncovariant) $\ell_\xi$. One can straightforwardly
demonstrate that the choice (\ref{lie1}) yields the following Noether
identities:
\begin{eqnarray}
D{\cal F}_\alpha &\equiv& e_\alpha\rfloor T^\beta\wedge{\cal F}_\beta
+ e_\alpha\rfloor R_\gamma{}^\beta\wedge{\cal F}^\gamma{}_\beta - {\frac 12}
\,(e_\alpha\rfloor Q_{\beta\gamma})f^{\beta\gamma}\nonumber\\
&&\qquad +\,(e_\alpha\rfloor D\psi^A)\wedge{\cal F}_A + (-1)^p(e_\alpha
\rfloor\psi^A)\wedge D{\cal F}_A,\label{NoeD1}\\
{\cal E}_\alpha &\equiv& e_\alpha\rfloor V^{\rm tot} + e_\alpha\rfloor
T^\beta\wedge {\cal H}_\beta + e_\alpha\rfloor R_\gamma{}^\beta\wedge
{\cal H}^\gamma{}_\beta + {\frac 12}\,(e_\alpha\rfloor Q_{\beta\gamma})
{\cal M}^{\beta\gamma}\nonumber\\
&&\qquad -\,(e_\alpha\rfloor D\psi^A)\wedge {\frac{\partial V^{\rm tot}}
{\partial D\psi^{A}}} - (e_\alpha\rfloor \psi^A)\wedge {\frac{\partial
V^{\rm tot}}{\partial \psi^{A}}}. \label{n02}
\end{eqnarray}

For the second choice (\ref{lie2}), in accordance with the above general
analysis, see (\ref{zero}), a zero term is added that is proportional to
\begin{equation}
N_\alpha{}^\beta := {\stackrel{\{\,\}}{\Gamma}}_\alpha{}^\beta
- \Gamma_\alpha{}^\beta.\label{K}
\end{equation}
This quantity is known as {\it distortion} 1-form. In particular,
the torsion is recovered from it as $T^\alpha = - N_\beta{}^\alpha
\wedge\vartheta^\beta$, whereas the nonmetricity arises as $Q_{\alpha\beta}
= - 2N_{(\alpha\beta)}$. It is straightforward to verify
the explicit formula
\begin{eqnarray}
N_{\alpha\beta} &=& -\,{\frac 12}\,Q_{\alpha\beta} + e_{[\alpha}
\rfloor n_{\beta]},\label{Nn}\\
n_\beta &:=& 2T_\beta - Q_{\beta\gamma}\wedge\vartheta^\gamma - {\frac 12}
\,e_\beta\rfloor\left(T_\gamma\wedge\vartheta^\gamma\right).\label{nb}
\end{eqnarray}
As we see, the symmetric part of the distorsion is determined by the
nonmetricity, while the skew-symmetric part is constructed from the
the 2-form $n_\alpha$. The latter has the property
\begin{equation}
n_\alpha\wedge\vartheta^\alpha =
\frac{1}{2}\,T_\alpha\wedge\vartheta^\alpha.\label{naxi}
\end{equation}
The corresponding (total and Riemannian) curvature 2-forms are related via
\begin{equation}\label{RR}
R_\alpha{}^\beta = {\stackrel{\{\,\}}{R}}_\alpha{}^\beta - {\stackrel
{\{\,\}}{D}}N_\alpha{}^\beta + N_\gamma{}^\beta\wedge N_\alpha{}^\gamma.
\end{equation}
Using these facts, we can verify that the diffeomorphism Noether identity
can be recast in the alternative form
\begin{eqnarray}
{\stackrel{\{\,\}}{D}}\left({\cal F}_\alpha - {\cal F}^\gamma{}_\beta
e_\alpha\rfloor N_\gamma{}^\beta\right) \equiv \left(e_\alpha\rfloor
{\stackrel{\{\,\}}{R}}_\gamma{}^\beta - {\stackrel{\{\,\}}
{\hbox{\L}}}_\alpha N_\gamma{}^\beta\right)\wedge
{\cal F}^\gamma{}_\beta\nonumber\\
+\,(e_\alpha\rfloor {\stackrel{\{\,\}}{D}}\psi^A)\wedge{\cal F}_A +
(-1)^p(e_\alpha\rfloor\psi^A)\wedge {\stackrel{\{\,\}}{D}}{\cal F}_A.
\label{NoeD2}
\end{eqnarray}
The identities (\ref{NoeD1}) and (\ref{NoeD2}) are equivalent, but in
contrast to the usual (\ref{NoeD1}), the alternative form (\ref{NoeD2})
is less known. The Lorentz-covariant version of this identity was derived
previously in \cite{PGrev} using a different method, see also \cite{invar}.
Although actually (\ref{NoeD2}) was used in \cite{NH} for the analysis of
the dynamics of test matter in MAG, the derivation of this identity is
published here for the first time.

\subsection{Yano derivative and the invariant conserved current for MAG}

Directly from the definition (\ref{lie3}), we can calculate the Yano
derivative for the covariant derivative of the matter field, of the torsion,
nonmetricity and curvature:
\begin{eqnarray}
{\cal L}_\xi D\psi^A &=& D({\cal L}_\xi \psi^A) + ({\cal L}_\xi\Gamma_\beta
{}^\alpha)\wedge (\rho^\beta{}_\alpha)^A_B\,\psi^B,\label{lieP}\\
{\cal L}_\xi T^\alpha &=&
D({\cal L}_\xi\vartheta^\alpha)+({\cal L}_\xi\Gamma_\beta
{}^\alpha)\wedge\vartheta^\beta, \label{lieT}\\
{\cal L}_\xi Q_{\alpha\beta} &=& - D({\cal L}_\xi g_{\alpha\beta})
+ ({\cal L}_\xi\Gamma_\alpha{}^\gamma)\,g_{\gamma\beta}
+ ({\cal L}_\xi\Gamma_\beta{}^\gamma)\,g_{\alpha\gamma},\label{lieQ}\\
{\cal L}_\xi R_\alpha{}^\beta &=& D({\cal L}_\xi
\Gamma_\alpha{}^\beta), \label{lieR}
\end{eqnarray}
Furthermore, it is straightforward to find the explicit expressions for
the Yano derivatives of the geometric and matter fields:
\begin{eqnarray}
{\cal L}_\xi\vartheta^\alpha &=& D\xi^\alpha + \xi\rfloor T^\alpha
- \Xi_\beta{}^\alpha\,\vartheta^\beta \equiv 0,\label{Ytheta}\\
{\cal L}_\xi\Gamma_\alpha{}^\beta &=& D\Xi_\alpha{}^\beta +
\xi\rfloor R_\alpha{}^\beta,\label{Ygamma}\\
{\cal L}_\xi g_{\alpha\beta} &=& -\xi\rfloor Q_{\alpha\beta} +
2\Xi_{(\alpha\beta)},\label{Yg}\\
{\cal L}_\xi\psi^A &=& D(\xi\rfloor\psi^A) + \xi\rfloor D\psi^A -
\Xi_\alpha{}^\beta\,(\rho^\alpha{}_\beta)^A_B\,\psi^B.\label{Ypsi}
\end{eqnarray}
Here we denote, as usual, $\xi^\alpha := \xi\rfloor\vartheta^\alpha$, and
\begin{equation}
\Xi_\alpha{}^\beta :=\xi\rfloor\Gamma_\alpha{}^\beta
+ \Theta_\alpha{}^\beta.\label{Xia}
\end{equation}
This is the MAG version of the general definition (\ref{XiG}).

Now, we replace in (\ref{vard2}) the ordinary Lie derivatives
$\ell_\xi$ with the Yano derivatives ${\cal L}_\xi$, make use of
(\ref{lieP})-(\ref{lieR}), and take into account the Noether identity
(\ref{Nlin2}). Then we can recast the identity (\ref{vard2}) as
\begin{equation}
{\cal L}_\xi\Gamma_\alpha{}^\beta\wedge {\cal F}^\alpha{}_\beta + {\frac 12}
\,{\cal L}_\xi g_{\alpha\beta}\wedge f^{\alpha\beta} + {\cal L}_\xi\psi^A
\wedge {\cal F}_A -\,d{\cal J}[\xi] = 0, \label{YlieV}
\end{equation}
where we introduced the scalar $(n-1)$-form
\begin{equation}\label{Jgrav1}
{\cal J}[\xi] := \xi\rfloor V^{\rm tot}
+ {\cal L}_\xi\Gamma_\alpha{}^\beta\wedge {\cal H}^\alpha{}_\beta
- {\frac 12}\,{\cal L}_\xi g_{\alpha\beta}\wedge {\cal M}^{\alpha\beta}
- {\cal L}_\xi\psi^A\wedge {\frac{\partial V^{\rm tot}}{\partial D\psi^{A}}}.
\end{equation}
Note that neither (\ref{YlieV}) nor (\ref{Jgrav1}) contain the terms
proportional to the Yano derivative of the coframe (cf. with \cite{invar})
because the latter is zero, cf. (\ref{Ytheta}).
By making use of the Yano derivative (\ref{Ytheta})-(\ref{Ypsi}), and taking
into account the Noether identities (\ref{n02}) and (\ref{Nlin2}), we recast
this current into the equivalent form
\begin{eqnarray}
{\cal J}[\xi] &=& d\left(\xi^\alpha\,{\cal H}_\alpha + \Xi_\alpha{}^\beta
{\cal H}^\alpha{}_\beta - \xi\rfloor\psi^A\wedge {\frac{\partial V^{\rm tot}}
{\partial D\psi^{A}}}\right)\nonumber\\
&&\qquad +\,\xi^\alpha\,{\cal F}_\alpha + \Xi_\alpha{}^\beta\,{\cal F}^\alpha
{}_\beta + \xi\rfloor\psi^A\wedge {\cal F}_A.\label{JMAG}
\end{eqnarray}

When the gravitational and matter variables satisfy the field equations
(\ref{onshellMAG}), we find that the current (\ref{Jgrav1}) is conserved,
$d{\cal J}[\xi] =0$, for any vector field $\xi$. By construction, this
conserved current is invariant under both diffeomorphism and local linear
transformations. Moreover, when (\ref{onshellMAG}) are satisfied, the
current  $(n-1)$-form is expressed in terms of a superpotential $(n-2)$-form,
(\ref{JMAG}). The corresponding conserved current then can be
calculated via the integral
\begin{equation}
{\cal Q}[\xi] = \int_{\partial S}\left(\xi^\alpha\,{\cal H}_\alpha +
\Xi_\alpha{}^\beta{\cal H}^\alpha{}_\beta - \xi\rfloor\psi^A\wedge
{\frac{\partial V^{\rm tot}}{\partial D\psi^{A}}}\right)\label{QMAG}
\end{equation}
over an $(n-2)$-dimensional boundary $\partial S$ of an
$(n-1)$-hypersurface $S$.

\section{Examples of MAG solutions}\label{solutions}

Let us now apply the general formalism to the exact solutions that
can be obtained with the help of the so called triplet ansatz in the
class of models with quadratic Lagrangians in $n = 4$ dimensions.
Such solutions are described, for instance, in
\cite{pono,triplet,plebdem,tw1,dot,tw2}.
The most general scheme for the triplet ansatz technique was developed
in \cite{triplet}, and the overview of the exact solutions in MAG models
can be found in \cite{magsol}.

\subsection{Gravitational field equations}

We consider the Lagrangian \cite{triplet} that generalizes the models
studied in \cite{pono,tw1,tw2,magexact,magkerr},
\begin{eqnarray}
V_{\rm MAG}&=&
\frac{1}{2\kappa}\,\left[-a_0\left(R^{\alpha\beta}\wedge\eta_{\alpha\beta}
-2\lambda\,\eta\right) +
T^\alpha\wedge{}^*\!\left(\sum_{I=1}^{3}a_{I}\,^{(I)}
T_\alpha\right)\right.\nonumber\\
&&+ 2\left(\sum_{I=2}^{4}c_{I}\,^{(I)}Q_{\alpha\beta}\right)
\wedge\vartheta^\alpha\wedge{}^*\!\, T^\beta + Q_{\alpha\beta}
\wedge{}^*\!\left(\sum_{I=1}^{4}b_{I}\,^{(I)}Q^{\alpha\beta}\right)
\nonumber\\
&&+ b_5({}^{(3)}Q_{\alpha\gamma}\wedge\vartheta^{\alpha})\wedge
{}^*({}^{(4)}Q^{\beta\gamma}\wedge\vartheta_{\beta})\Bigg] - \frac{1}{2}
z_4\,R^{\alpha\beta}\wedge{}^*{}^{(4)}Z_{\alpha\beta}.\label{lagr}
\end{eqnarray}
Here, the coupling constants $a_0,...,a_3,c_2,c_3,c_4, b_1,...,b_5, z_4$ are
dimensionless, $\kappa := 8\pi G/c^3$ is the standard Einstein gravitational
constant, and $\lambda$ is the cosmological constant. The segmental curvature
is denoted by ${}^{(4)}Z_{\alpha\beta}:={1\over 4}g_{\alpha\beta}R_{\gamma}
{}^{\gamma}$; it is a purely {\it post}-Riemannian piece.

This Lagrangian is constructed from the irreducible parts of the
torsion and nonmetricity. Namely, let us recall that the torsion 2-form
can be decomposed into three irreducible pieces, $T^{\alpha}={}^{(1)}
T^{\alpha} + {}^{(2)}T^{\alpha} + {}^{(3)}T^{\alpha}$, where
\begin{eqnarray}
{}^{(2)}T^{\alpha}&:=&{1\over 3}\vartheta^{\alpha}\wedge T,\label{T2}\\
{}^{(3)}T^{\alpha}&:=&-\,{1\over 3}{}^*(\vartheta^{\alpha}\wedge P),
\label{T3}\\
{}^{(1)}T^{\alpha}&:=&T^{\alpha}-{}^{(2)}T^{\alpha} - {}^{(3)}T^{\alpha}.
\label{T1}
\end{eqnarray}
The torsion trace (covector) and pseudotrace (axial covector) 1-forms are
defined, respectively, by
\begin{equation}
T:=e_{\alpha}\rfloor T^{\alpha},\quad\quad
 P:={}^*(T^{\alpha}\wedge\vartheta_{\alpha}).\label{TP}
\end{equation}
Analogously, the nonmetricity 1-form can be decomposed into four
irreducible pieces, $Q_{\alpha\beta}={}^{(1)}Q_{\alpha\beta}+{}^{(2)}
Q_{\alpha\beta}+ {}^{(3)}Q_{\alpha\beta}+{}^{(4)}Q_{\alpha\beta}$, with
\begin{eqnarray}
{}^{(2)}Q_{\alpha\beta}&:=&{2\over 3}\,{}^*\!\left[\vartheta_{(\alpha}\wedge
\Omega_{\beta)}\right],\label{Q2}\\
{}^{(3)}Q_{\alpha\beta}&:=&{4\over 9}
\left[\vartheta_{(\alpha}e_{\beta)}
\rfloor\Lambda - {1\over 4}g_{\alpha\beta}\Lambda\right],\label{Q3}\\
{}^{(4)}Q_{\alpha\beta}&:=&g_{\alpha\beta}Q,\label{Q4}\\
{}^{(1)}Q_{\alpha\beta}&:=&Q_{\alpha\beta}-{}^{(2)}Q_{\alpha\beta}-
{}^{(3)}Q_{\alpha\beta}-{}^{(4)}Q_{\alpha\beta}.\label{Q1}
\end{eqnarray}
Here the shear covector part and the Weyl covector are, respectively,
\begin{equation}
\Lambda:=\vartheta^{\alpha}e^{\beta}\rfloor
{\nearrow\!\!\!\!\!\!\!Q}_{\alpha\beta},\quad\quad
Q:={1\over 4}g^{\alpha\beta}Q_{\alpha\beta},\label{LQ}
\end{equation}
where ${\nearrow\!\!\!\!\!\!\!Q}_{\alpha\beta}=Q_{\alpha\beta}-
Qg_{\alpha\beta}$ is the traceless piece of the nonmetricity. The 2-form
$\Omega^{\alpha}$ is defined by
$\Omega_{\alpha}:=\Theta_{\alpha} - {1\over 3}e_{\alpha}\rfloor
(\vartheta^{\beta}\wedge\Theta_{\beta})$ with $\Theta_{\alpha}:=
{}^*({\nearrow\!\!\!\!\!\!\!Q}_{\alpha\beta}\wedge\vartheta^{\beta})$.
We can prove that $e_{\alpha}\rfloor\Omega^{\alpha}=0$ and
$\vartheta_{\alpha}\wedge\Omega^{\alpha}=0$.

In order to write down the field equations (\ref{first}), (\ref{second}),
(\ref{onshellMAG}), we need the field momenta (\ref{HHM}) and the
generalised potentials (\ref{EEm}). A straightforward computation yields
for the Lagrangian (\ref{lagr}):
\begin{eqnarray}
{\cal M}^{\alpha\beta} &=&
-{2\over\kappa}\Bigg[{}^*\! \left(\sum_{I=1}^{4}b_{I}{}^{(I)}
Q^{\alpha\beta}\right) + {1\over 2}b_5\left(\vartheta^{(\alpha}\wedge{}^*
(Q\wedge\vartheta^{\beta)}) - {1\over 4}g^{\alpha\beta}\,{}^*(3Q + \Lambda)
\right)\nonumber\\
&& +\, c_{2}\,\vartheta^{(\alpha}\wedge{}^*\! ^{(1)}T^{\beta)} +
c_{3}\,\vartheta^{(\alpha}\wedge{}^*\! ^{(2)}T^{\beta)} +
{1\over 4}(c_{3}-c_{4})\,g^{\alpha\beta}{}^*\!\,  T\Bigg]\,,\label{M1}\\
{\cal H}_{\alpha} &=&
- {1\over\kappa}\,
  {}^*\!\left[\left(\sum_{I=1}^{3}a_{I}{}^{(I)}T_{\alpha}\right) +
    \left(\sum_{I=2}^{4}c_{I}{}^{(I)}
      Q_{\alpha\beta}\wedge\vartheta^{\beta}\right)\right],\label{Ha1}\\
{\cal H}^{\alpha}{}_{\beta} &=&
{a_0\over 2\kappa}\,\eta^{\alpha}{}_{\beta} +
  z_4\,{}^{*}\!\left({}^{(4)}Z^{\alpha}{}_{\beta}\right),\label{Hab1}
\end{eqnarray}
and, in accordance with the Noether identities (\ref{Nlin2}) and (\ref{n02}),
\begin{eqnarray}
{\cal E}_{\alpha} &=& e_{\alpha}\rfloor V_{\rm MAG}
+ (e_{\alpha}\rfloor T^{\beta})\wedge{\cal H}_{\beta} + (e_{\alpha}\rfloor
R_{\beta}{}^{\gamma})\wedge{\cal H}^{\beta}{}_{\gamma} + {1\over 2}(
e_{\alpha}\rfloor Q_{\beta\gamma}){\cal M}^{\beta\gamma},\label{Ea}\\
{\cal E}^{\alpha}{}_{\beta} &=& - \vartheta^{\alpha}\wedge {\cal H}_{\beta}
- {\cal M}^{\alpha}{}_{\beta}.\label{Eab}
\end{eqnarray}
As we already mentioned, the equation which arises from the variation
of the Lagrangian with respect to the metric turns out to be redundant.

\subsection{Triplet ansatz and effective system}

The triplet ansatz specifies a particular structure for the
post-Riemannian sector of MAG model. Namely, it is assumed that this
sector is totally described by the three covectors: the 1-form of the
torsion trace $T$, the Weyl 1-form $Q$, and the nonmetricity 1-form
$\Lambda$, defined in (\ref{TP}) and (\ref{LQ}). Moreover, they are
all proportional to an auxiliary 1-form:
\begin{equation}
Q = k_0\,A_{\rm MAG},\qquad \Lambda = k_1\,A_{\rm MAG},\qquad
T = k_2\,A_{\rm MAG},\label{tri}
\end{equation}
with constant coefficients $k_0, k_1, k_2$; the 1-form $A_{\rm MAG}$
is a new variable to be determined from the field equations. Accordingly,
${}^{(1)}T^{\alpha} = {}^{(3)}T^{\alpha} =0$ and ${}^{(1)}Q_{\alpha\beta}
= {}^{(2)}Q_{\alpha\beta} =0$.

Substituting the triplet ansatz into the MAG field equations, we find
the explicit form of the above coefficients in terms of the coupling
constants of the MAG Lagrangian:
\begin{eqnarray}
k_0 &=& \left({\frac {a_2}2} - a_0\right)(8b_3 + a_0)
- 3(c_3 + a_0)^2,\label{k0}\\
k_1 &=& -\,9\left[(a_0 - b_5)\left({\frac {a_2}2} - a_0\right)
+ (c_3 + a_0)(c_4 + a_0)\right],\label{k1}\\
k_2 &=& {\frac 32}\left[3(a_0 - b_5)(c_3 + a_0) + (8b_3 + a_0)
(c_4 + a_0)\right],\label{k2}
\end{eqnarray}
and in addition,
\begin{equation}
- 4k_0b_4 + {\frac {k_1}2}\,b_3 + k_2c_4 + {\frac {a_0}2}k =0,\label{b4}
\end{equation}
with $k:=3k_0-k_1+2k_2$.

As a result, the MAG field equations (\ref{first}), (\ref{second}) for the
Lagrangian (\ref{lagr}) reduce to the effective Einstein-Maxwell system:
\begin{eqnarray}
{\frac 1 2}\,\widetilde{R}^{\mu\nu}\wedge\eta_{\alpha\mu\nu}
- \lambda\,\eta_\alpha&=&\kappa\,\Sigma^{\rm MAG}_\alpha,\label{effE}\\
d\,{}^\ast\!F_{\rm MAG} &=& 0.\label{effM}
\end{eqnarray}
Here $F_{\rm MAG} = dA_{\rm MAG}$, and the effective energy-momentum reads
\begin{equation}
\Sigma^{\rm MAG}_\alpha = {\frac { Y_{\rm MAG}}2}\left[F_{\rm MAG}\wedge
\left(e_\alpha\rfloor{}^\ast F_{\rm MAG}\right) - {}^\ast F_{\rm MAG}\wedge
\left(e_\alpha\rfloor F_{\rm MAG}\right)\right].
\end{equation}
The effective ``vacuum constant" is defined by $ Y_{\rm MAG} :=
-z_4 k_0^2/a_0$, and the tilde as usual denotes the objects constructed
from the Riemannian (Christoffel) connection.

\subsection{Axially symmetric solution}

Let us choose standard Boyer-Lindquist coordinates $(t,r,\theta,\varphi)$.
Then we straightforwardly can verify that the MAG field equations admit
the generalized Kerr-Newman solution with cosmological constant. It is
described by the coframe
\begin{eqnarray}
\vartheta ^{\hat{0}} &=&\,\sqrt{{\Delta}\over{\Sigma}}\,\left(\,cdt
-j_0\Omega\sin ^2\theta\,d\varphi\,\right),\label{cof0}\\
\vartheta ^{\hat{1}} &=&\,\sqrt{{\Sigma}\over{\Delta}}\;d\,r,\label{cof1}\\
\vartheta ^{\hat{2}} &=&\,\sqrt{{\Sigma}\over f}\;d\,\theta,\label{cof2}\\
\vartheta ^{\hat{3}} &=&\,\sqrt{f\over{\Sigma}}\,\sin\theta\,\left[
\,- j_0 cdt  +\Omega\left(\,r^2+j_0^2\,\right)
d\varphi\,\right],\label{cof3}
\end{eqnarray}
and by the 1-form that describes post-Riemannian triplet sector
\begin{equation}
A_{\rm MAG} = u\,\vartheta^{\hat{0}},\label{Amag}
\end{equation}
Here $\Delta=\Delta(r), \Sigma=\Sigma(r,\theta), f=f(\theta), u = u(r,
\theta)$, and
 $j_0$ and $\Omega$ are constant. The field equations yield the
following explicit functions
\begin{eqnarray}
\Delta &=& (r^2 + j_0^2)\left(1 - {\frac \lambda 3}\,r^2\right) - 2mr
+{\frac {\kappa Y_{\rm MAG}}2}\,N^2,\label{sol2a}\\
\Sigma &=& \,r^2 +j_0^2\cos^2\theta,\label{sol2b}\\
f &=&\,1+{\frac \lambda 3}\,j_0^2\cos ^2\theta\,,\label{sol2c}\\
u &=& {\frac {N\,r}{\sqrt{\Delta\Sigma}}}.\label{sol2d}
\end{eqnarray}
Here $\Omega = (1 + \lambda j_0^2/3)^{-1}$, $m = GM/c^2$, with the
arbitrary integration constants $M$ and $N$.

For completeness, let us write down explicitly the post-Riemannian 2-form
\begin{equation}
F_{\rm MAG} = N\,d\left({r\over\sqrt{\Delta\Sigma}}\,\vartheta^{\hat{0}}
\right) ={\frac N {\Sigma^2}}\left[(r^2 - j_{0}^2\cos^2\theta)
\,\vartheta^{\hat{0}}\wedge\vartheta^{\hat{1}} - 2j_{0}r\cos\theta
\,\vartheta^{\hat{2}}\wedge\vartheta^{\hat{3}}\right].\label{Fmag}
\end{equation}

\subsection{Invariant conserved charges for the axially symmetric solution}

In order to calculate the conserved charge (\ref{QMAG}), we need the
explicit translational and linear field momenta. Using the triplet ansatz
in (\ref{Ha1}) and (\ref{Hab1}), we find
\begin{eqnarray}
{\cal H}_\alpha &=& -\,{\frac {a_0k}{3\kappa}}\,{}^*\left(\vartheta_\alpha
\wedge A_{\rm MAG}\right) = {\frac {a_0k}{3\kappa}}\,e_\alpha\rfloor
{}^*A_{\rm MAG},\label{Ha2}\\
{\cal H}^\alpha{}_\beta &=& {a_0\over 2\kappa}\,\eta^{\alpha}{}_{\beta} +
{\frac {z_4} 8}\,\delta^\alpha_\beta\,{}^*\!dQ = {a_0\over 2\kappa}
\,\eta^{\alpha}{}_{\beta} + {\frac {z_4k_0} 8}\,\delta^\alpha_\beta
\,{}^*\!F_{\rm MAG}.\label{Hab2}
\end{eqnarray}
{}From (\ref{Ytheta}) we have explicitly
$\Xi_\beta{}^\alpha = e_\beta\rfloor\left(D\xi^\alpha
+ \xi\rfloor T^\alpha\right).$
However, it is more convenient to start directly from the definition
(\ref{Xia}), and substitute (\ref{K}) in it. We then find
 \begin{equation}
\Xi_\beta{}^\alpha = {\stackrel{\{\,\}}{D}}_\beta\xi^\alpha
- \xi\rfloor N_\beta{}^\alpha.\label{Xiab2}
\end{equation}
By making use of (\ref{Nn}), we then can easily compute the contractions:
\begin{eqnarray}
\Xi_\alpha{}^\alpha &=& {\stackrel{\{\,\}}{D}}_\alpha\xi^\alpha
+ 2\xi\rfloor Q,\label{Xiaa1}\\
\eta^{\alpha\beta}\,\Xi_{\alpha\beta} &=& {}^*\!\left[\,d(\xi_\alpha
\vartheta^\alpha) - \xi^\alpha n_\alpha + \frac{1}{2}\,\xi\rfloor
(T_\alpha\wedge
\vartheta^\alpha)\right].\label{etaXi1}
\end{eqnarray}
Furthermore, in the triplet ansatz framework, these formulas reduce to
\begin{eqnarray}
\Xi_\alpha{}^\alpha &=& {\stackrel{\{\,\}}{D}}_\alpha\xi^\alpha
+ 2k_0\,\xi\rfloor A_{\rm MAG},\label{Xiaa2}\\
\eta^{\alpha\beta}\,\Xi_{\alpha\beta} &=& {}^*d(\xi_\alpha\vartheta^\alpha)
+ {\frac {k}3}\,\xi\rfloor{}^*A_{\rm MAG}.\label{etaXi2}
\end{eqnarray}

Now we are in a position to compute the conserved invariant charges.
Substituting (\ref{Ha2}), (\ref{Hab2}), (\ref{Xiaa2}) and (\ref{etaXi2})
into (\ref{QMAG}), we find the general expression for the invariant
charge in the quadratic MAG model with a triplet ansatz:
\begin{eqnarray}
{\cal Q}[\xi] &=& {\frac {a_0}{2\kappa}}\int_{\partial S}\,\left[{}^*\,d(
\xi^\alpha\vartheta_\alpha) + k\,(\xi\rfloor{}^*A_{\rm MAG})\right]\nonumber\\
&& +\,{\frac {z_4k_0}{8}}\int_{\partial S}\,({\stackrel
{\{\,\}}{D}}_\alpha\xi^\alpha + 2k_0\,\xi\rfloor A_{\rm MAG})
\,{}^*F_{\rm MAG}.\label{Qsol}
\end{eqnarray}

On the axially symmetric solution (\ref{cof0})-(\ref{sol2d}), the last
integral is zero for $\xi = \xi^i\partial_i$ with the constant components
$\xi^i$. The asymptotic behavior of the physical and geometrical
quantities is as follows (keeping the leading terms in $1/r$):
\begin{eqnarray}
\xi\rfloor A_{\rm MAG} &=& {\frac Nr}\left(\xi^0 - \xi^3\,\Omega 
j_0\sin^2\theta\right),\label{Aasy}\\
{}^*A_{\rm MAG} &=& {\frac {Nr}{\Delta}}\,dr\wedge\sin\theta d\theta
\wedge\left(-j_0cdt + \Omega(r^2 + j_0^2)d\varphi\right),\label{astA}\\
{}^*F_{\rm MAG} &=& -\,N\Omega\sin\theta d\theta\wedge d\varphi.\label{Fasy}
\end{eqnarray}
Substituting this, we can verify that the second line in (\ref{Qsol}) vanishes,
and for $\lambda = 0$ the invariant conserved charges read
\begin{equation}
{\cal Q}[\partial_t] = a_0Mc^2/2,\qquad {\cal Q}[\partial_\varphi] 
= - a_0Mcj_0.\label{Qex} 
\end{equation}
When the coupling constant in the MAG Lagrangian (\ref{lagr}) has its 
standard value $a_0 = 1$, these conserved charges reduce to the well known
results of Komar \cite{invar}. Note that $a_0$ cannot vanish, otherwise the
quadratic MAG model does not have an Einsteinian limit and the triplet 
ansatz is not applicable. For the solutions with nontrivial cosmological
constant $\lambda \neq 0$, the conserved charges are formally infinite.
Accordingly, like in the case of the theories with local Lorentz symmetry, 
a regularization is required. Unfortunately, the usual regularization via
relocalization with the help of the topological boundary term is not possible
since in MAG there is no analogue \cite{CS} of the Euler invariant. We will 
analyze the regularization problem elsewhere.

\section{Discussion and conclusion}

In this paper, the approach, developed earlier in \cite{invar}, is 
generalized to the case when the local Lorentz group is replaced by an 
arbitrary local gauge group. The scheme includes the Maxwell and Yang-Mills 
fields coupled to gravity with Abelian and non-Abelian local internal 
symmetries. However, in the case of internal symmetries there seems to be no 
natural way to define nontrivial covariant Lie derivatives. In the 
gravitational case, on the other hand, the frame field, transforming 
covariantly under the symmetry group, is the key ingredient to construct 
covariant generalized (Yano) Lie derivatives and therefore, invariant 
conserved quantities. The existence of the frame field in gravitational 
theories can be understood in terms of the so-called soldering procedure.  
In Sec.~\ref{internal} we have studied a particular case in which a scalar 
field plays a role analogous to the frame field for the external symmetries, 
allowing to construct $U(1)$-invariant conserved quantities.

Another important case is the metric-affine gravity in which the local Lorentz 
spacetime group is extended to the local general linear group. We have developed
the corresponding general formalism for MAG in Sec.~\ref{external}. This scheme
generalizes and refines the partial results available in the earlier literature
\cite{Fer94,Bor94,sard1,sard2,sard3,Mielke01,Hecht92}. In order to illustrate
how the formalism works, we applied it to the computation of the conserved charges 
for an exact MAG solution in Sec.~\ref{solutions}. The results obtained are 
consistent with the derivations for the general relativity and for the models 
with local Lorentz symmetry.

\bigskip
{\bf Acknowledgments}. We thank J.G. Pereira for his hospitality at IFT-UNESP 
where this work was started. This work was partially supported by FAPESP and 
by DFG, project He~528/21-1 (for YNO) and by CNPq and by FONDECYT grant 
\#~1060939 (for GFR). We also thank F.W. Hehl for his comments on our paper.

\appendix
\section{Generalized Lie derivatives}\label{app1}

Here, we collect some useful identities satisfied by the generalized Lie
derivative ${\cal L}_{\xi,\varepsilon}$. They generalize the results found in
Appendix A of \cite{invar}.

We defined the generalized Lie derivative by the formula
\begin{equation}
{\cal L}_{\xi,\varepsilon}\omega^A := \ell_\xi\omega^A -
\varepsilon^a(\rho_a)^A{}_B\,\omega^B\label{genLie}
\end{equation}
when acting on any (gauge-)covariant $p$-form $\omega^A$.
We also define the generalized Lie derivative of the gauge field by
\begin{equation}
{\cal L}_{\xi,\varepsilon} A^a := \ell_\xi A^a
-d\varepsilon^a+f^a{}_{bc}\,A^b\varepsilon^c.\label{alt01}
\end{equation}

With these definitions, one can prove that the generalized Lie
derivative commutes with the exterior derivative, i.e. $\left[{\cal
L}_{\xi,\varepsilon}, d\right]=0$, as well as the following identities:
\begin{eqnarray}
{\cal L}_{\xi,\varepsilon} A^a &\equiv& D\left(\xi\rfloor
A^a-\varepsilon^a\right)+\xi\rfloor F^a,\\
{\cal L}_{\xi,\varepsilon} F^a &\equiv&
D({\cal L}_{\xi,\varepsilon}A^a), \\
\left[{\cal L}_{\xi,\varepsilon}, D\right]\omega^A&\equiv&
({\cal L}_{\xi,\varepsilon}A^a)(\rho_a)^A{}_B\wedge\omega^B.
\end{eqnarray}

\end{document}